\documentclass[10pt,aps,prb,groupedaddress,amsmath,amssymb]{revtex4}
\usepackage{times}
\usepackage{amsmath}
\usepackage{dcolumn}
\usepackage{graphicx}
\usepackage{color}
\usepackage{ulem}
\usepackage{subfigure}
\def\OpOne{\hat{\mathbf{1}}}
\def\Hhat{\hat{H}}

\begin{document}

\title{Spectroscopic accuracy directly from quantum chemistry: application to ground and excited states of beryllium dimer}

\author{Sandeep Sharma}
\affiliation{  Department of Chemistry, Frick Laboratory, Princeton University, NJ 08544}

\author{Takeshi Yanai}
\affiliation{Department of Theoretical and Computational Molecular Science, Institute for Molecular Science, Okazaki, Aichi 444-8585, Japan}

\author{George H. Booth}
\affiliation{~Department of Chemistry, Frick Laboratory, Princeton University, NJ 08544}

\author{C. J. Umrigar}
\affiliation{Laboratory of Atomic and Solid State Physics, Cornell University, NY 14853}

\author{Garnet Kin-Lic Chan*}
\affiliation{Department of Chemistry, Frick Laboratory, Princeton University, NJ 08544}
\email{gkc1000@gmail.com}

\begin{abstract}
We combine explicit correlation via the canonical transcorrelation approach with the density matrix renormalization group
and initiator full configuration interaction quantum Monte Carlo methods to compute
a near-exact beryllium dimer curve, {\it without} the
use of composite methods.  In particular, our direct density matrix renormalization group calculations produce a well-depth of $D_e$=931.2 cm$^{-1}$ which
agrees very well with recent experimentally derived estimates
$D_e$=929.7$\pm 2$~cm$^{-1}$ [Science, {\bf 324}, 1548 (2009)] and $D_e$=934.6~cm$^{-1}$ [Science, {\bf 326}, 1382 (2009)]],
as well the best composite theoretical estimates,
$D_e$=938$\pm 15$~cm$^{-1}$ [J. Phys. Chem. A, {\bf 111}, 12822 (2007)] and $D_e$=935.1$\pm 10$~cm$^{-1}$ [Phys. Chem. Chem. Phys., {\bf 13}, 20311 (2011)]. 
Our results suggest possible inaccuracies in the functional form of the potential used at shorter bond lengths to fit the experimental data [Science, {\bf 324}, 1548 (2009)]. With the density matrix renormalization group
we also compute near-exact vertical 
excitation energies at the equilibrium geometry. These
provide non-trivial benchmarks for quantum chemical methods for excited states, and illustrate the surprisingly large error
that remains for 1$^1\Sigma^-_g$ state with approximate multi-reference configuration interaction and equation-of-motion coupled cluster methods.  Overall, we demonstrate that explicitly correlated
density matrix renormalization group and initiator full configuration interaction quantum Monte Carlo methods allow us to fully
converge to the basis set and correlation limit of the non-relativistic Schr\"odinger equation in small molecules.
\end{abstract}

\maketitle

\section{Introduction}
In basis-set quantum chemistry, we divide the challenge of solving the non-relativistic electronic Schr\"odinger equation into
two parts: the treatment of $n$-electron correlations, and the saturation of the one-particle basis.
Recent years have seen significant advances in both these areas. In the first case, methods
such as general order coupled cluster (CC) \cite{kallay}, the density matrix renormalization group (DMRG) \cite{chan2011}, and
initiator full configuration interaction quantum Monte Carlo (\mbox{{\it i}-FCIQMC}) \cite{cleland,booth1} have been developed to
achieve an efficient treatment of arbitrary $n$-electron correlations in modestly sized molecules. In the second case, explicit correlation (F12) techniques\cite{Ten-no2004}
augment the one-particle basis with geminal functions that represent the electron-electron cusp.
Taken together, these advances provide the potential to converge to near-exact solutions of the non-relativistic electronic Schr\"odinger equation
at the basis set limit.
In this report, we describe the efficient combination of explicit correlation, via the canonical transcorrelation approach \cite{yanai2012canonical},
with DMRG and with \mbox{{\it i}-FCIQMC}, and apply these combinations to determine the ground and excited state
electronic structure of the beryllium dimer {to very high} accuracy.

\section{Methods}
The essence of explicit correlation (henceforth referred to as F12 theory) is to use a geminal correlation
factor, $f(r_{12})=-\frac{1}{\gamma}\exp(-\gamma r_{12})$ to augment the doubles manifold of the virtual space \cite{Kong2011,tennoreview}.
The geminal can be thought of as including some excitations into a formally infinite basis of virtuals. Labelling the
infinite virtual basis by $\alpha, \beta, \gamma, \ldots$, the geminal doubles excitation operator is written as
\begin{align}
T^{F_{12}} = \sum_{ij\alpha \beta} G^{ij}_{\alpha \beta} E_{ij}^{\alpha\beta} \label{eq:geminal_excite}
\end{align}
where $G^{ij}_{\alpha\beta}$ are the geminal doubles amplitudes.
A significant practical advance was the realization that the geminal
amplitudes are fixed to linear order by the electron-electron cusp condition \cite{Kutzelnigg1985,kut_klopp,Ten-no2004},
\begin{align}
G^{ij}_{\alpha\beta}= \frac{3}{8}\langle \alpha\beta|{Q}_{12} f(r_{12})|ij\rangle + \frac{1}{8} \langle \alpha\beta|{Q}_{12}f(r_{12})|ji\rangle
\end{align}
where ${Q}_{12}$ is a projector, defined in terms of projectors ${O}$ and ${V}$ into
the occupied and virtual space of the standard orbital basis,
\begin{align}
{Q}_{12} = (1-{O}_1)(1-{O}_2) - {V}_1 {V}_2
\end{align}
that ensures that excitations of the geminal factor are orthogonal to those of the standard orbital space \cite{sinanoglu}.

Combining F12 methodology with the DMRG and \mbox{{\it i}-FCIQMC} methods involves practical hurdles not
present in prior combinations of F12 theory with other correlation methods.
For example, explicitly correlated coupled cluster theory formally starts from
a non-Hermitian effective Hamiltonian, obtained by similarity transforming with the geminal excitation operator $\exp(T^{F_{12}})$ \cite{torheyden}
but the DMRG is most conveniently implemented with a Hermitian effective Hamiltonian. Similarly, the
 universal perturbative correction of Torheyden and Valeev, which has been used with \mbox{{\it i}-FCIQMC} \cite{booth,torheyden2} does not introduce non-Hermiticity, but 
requires
the one- and two-particle reduced density matrices which can be expensive to compute precisely in Monte Carlo methods.
These practical complications are removed within  the recently introduced canonical transcorrelation
form of F12 theory of Yanai and Shiozaki \cite{yanai2012canonical}. In this method,
a Hermitian effective Hamiltonian is obtained from
an {\it anti-hermitian} geminal doubles excitation operator, ${A}^{F_{12}}$
\begin{align}
{A}^{F_{12}} = \frac{1}{2} ({T}^{F_{12}} - {T}^{{F_{12}\dag}})
\end{align}
The canonical transcorrelated Hamiltonian is formally defined as
\begin{align}
\bar{H} = \exp({-{A}^{F^{12}}}) H \exp({{A}^{F_{12}}}) \label{eq:ct_h}
\end{align}
The fully transformed Hamiltonian involves operators
of high particle rank. To ameliorate the complexity, Yanai and Shiozaki invoke the same
commutator approximations used in the canonical transformation theory \cite{Neuscamman2010, yanaict1, yanaict2}, and further simplify the
quadratic commutator term by replacing the
Hamiltonian with a generalized Fock operator ${f}$,
\begin{align}
\bar{H}^{F_{12}} = H + [H, A^{F_{12}}]_{1,2} + \frac{1}{2}[[f, A^{F_{12}}], A^{F_{12}}]_{1,2} \label{eq:ct_f12}
\end{align}
an approximation which is valid through second-order in perturbation theory. The subscript $1,2$ denotes
that only one- and two-particle rank operators and density matrices are kept in the Mukherjee-Kutzelnigg normal-ordered form \cite{Mukherjee1997,kutzelnigg97}.
In our calculations here, normal-ordering is carried out with respect to the Hartree-Fock reference, thus no density 
cumulants \cite{Mazziottiijqc,kutzelniggcumulant,Mazziotticpl,MazziottiReview} appear. The only error arises from the neglect of the three-particle normal-ordered operator\cite{EricQuadraticCT} generated by the $A^{F_{12}}$ excitations. As the orbital basis increases, $A^{F_{12}}$ tends to zero and the three-particle error also goes to zero, very 
different behaviour from density cumulant theories where full three-particle quantity reconstruction is performed \cite{EricQuadraticCT,Mazziottijcp}. In this sense, 
the three-particle error in this theory is part of the basis set error.

The practical advantages of the canonical transcorrelation formulation are that no correlated density matrices are required
and the effective Hamiltonian is Hermitian and of two-particle form. It may thus be combined readily
with {\it any correlation treatment}.
Beyond the practical advantages, the canonical transcorrelation formulation uses a
``perturb then diagonalize'' approach, rather than the ``diagonalize then perturb'' approach of the a posteriori F12 treatment of Valeev previously combined with \mbox{{\it i}-FCIQMC}. This allows
the geminal factors to automatically relax the parameters of the subsequent correlation treatment. This approach is similar in spirit to the similarity transformed F12 method of Ten-no which has been used with the PMC-SD method of Ohtsuka \cite{ohtsuka}, but there the excitation operator is not anti-hermitian and alternative approximations are used in the simplification of the resulting equations. The use of the effective Hamiltonian (\ref{eq:ct_f12}) was denoted by Yanai and Shiozaki by the prefix
F12-, thus in their nomenclature, the combinations with DMRG and \mbox{{\it i}-FCIQMC} in this work would be \mbox{F12-DMRG} and \mbox{F12-{\it i}-FCIQMC} respectively. However,
as all our DMRG and \mbox{{\it i}-FCIQMC} calculations use this effective Hamiltonian here, we will usually omit the F12 prefix and simply refer to DMRG and \mbox{{\it i}-FCIQMC}.

We now briefly introduce the DMRG and \mbox{{\it i}-FCIQMC} correlation methods used in this work. The DMRG is
a variational ansatz based on a matrix-product representation of the FCI amplitudes.
Expanding the FCI wavefunction as
\begin{align}
|\Psi\rangle = \sum_{\{n\}} C^{n_1 n_2 \ldots n_k}|n_1 n_2 \ldots n_k\rangle
\end{align}
where $n_i$ is the occupancy of orbital $i$ in the occupancy vector representation of the $n$-particle determinant $|n_1 n_2 \ldots n_k\rangle$,
 $\sum_i n_i = n$, the one-site DMRG wavefunction approximates the FCI coefficient $C^{n_1 n_2 \ldots n_k}$
as the vector, matrix, \ldots, matrix, vector product
\begin{align}
C^{n_1 n_2 \ldots n_k} = \sum_{\{i \}}  A^{n_1}_{i_1} A^{n_2}_{i_1 i_2} \ldots A^{n_k}_{i_{k-1}}  \label{eq:mps}
\end{align}
For each occupancy $n_i$, the dimension of the corresponding matrix(vector) is  $M\times M$($M$). $M$ is usually referred to
as the number of renormalised states. The energy is determined  by minimizing
$\langle \Psi | H |\Psi\rangle / \langle \Psi|\Psi\rangle$ with respect to the matrix and vector coefficients in Eq. (\ref{eq:mps}) \cite{Chan2002,Hachmann2006,chan2011,marti11}.
As $M$ is increased, the DMRG energy converges towards the FCI limit. In practical DMRG calculations, the minimization is
carried out with a slightly more flexible wavefunction form, where two  $A^{n_r}$, $A^{n_{r}+1}$ matrices on adjacent orbitals are fused into a single larger composite (two-site) matrix, $A^{n_rn_{r+1}}$.
This introduces a larger variational space than in Eq. (\ref{eq:mps}), which improves the numerical convergence.
A measure of the error in a DMRG calculation is provided by the ``discarded'' weight, which is the (squared) difference in overlap between the two-site wavefunction
and its best one-site approximation: this difference vanishes as $M\to \infty$. The discarded weight usually exhibits
a linear relationship with the energy, and thus provides a convenient way to extrapolate the energy to the exact FCI result \cite{legeza,Chan2002}.

The FCIQMC algorithm has been recently introduced by Alavi and co-workers
\cite{Booth2013,booth1,cleland,semistochastic}.
FCIQMC is a projector Monte Carlo method
wherein the stochastic walk is done in determinant space~\cite{BlaSug-PRD-83,trivedi},
but instead of imposing a fixed node approximation~\cite{Haaf1995} it uses computational
power and cancellation algorithms to control the fermion sign problem.
The exact ground state wavefunction is obtained by repeatedly applying a ``projector" to an initial state,
\begin{align}
|\Psi\rangle = \lim_{n \to \infty} \left(\OpOne + \tau (E\OpOne-\Hhat)\right)^n |\Phi\rangle
\end{align}
If $|\Psi\rangle$ is expanded in an orthogonal basis of $N_s$ determinants, $|\Psi\rangle= \sum_{i=1}^{N_s} c_i |\mathbf{n}_i\rangle$
the expansion coefficients evolve according to
\begin{align}
c_i(t+1) = \left(1 + \tau(E-H_{ii})\right) c_i(t)  - \tau \sum_{j\neq i}^{N_s} H_{ij} c_j \label{eq:pop_dynamics}
\end{align}
where $t$ labels the iterations, and $\tau$ is a time step, the maximum value of which is constrained
by the inverse of the spectral range of the Hamiltonian.
Since the number of basis states, $N_s$ is too large to permit storing all the coefficients, $c_j$, a stochastic
approach is used wherein $N_w$ ``walkers" ($N_w \ll N_s$) sample the wavefunction.
Although the distribution of walkers among the states at any time step $t$ is a crude approximation to the
wavefunction, the infinite time average yields the ground state wavefunction exactly.
The term $1+\tau(E-H_{ii})$ in Eq.~(\ref{eq:pop_dynamics}) leads to an increase or decrease in the weight of the walker on determinant $i$ while $-\tau H_{ij}$ causes transitions of
walkers from determinant $j$ to determinant $i$. If walkers land on the same determinant, their weights are combined.
However, because contributions to a given determinant can be of either sign for most systems,
a fermionic sign problem results, where the signal becomes exponentially small compared to the noise\cite{foulkes}.
As demonstrated by Alavi and coworkers, however, when cancellations are employed, for {\it sufficiently large} $N_w$, the walk undergoes a transition into a regime where the sign problem is controlled\cite{booth1}.

For the sufficiently large $N_w$ such that this cancellation is effective, FCIQMC is exact within a statistical error of order $\sim (N_w N_t)^{-1/2}$. However, the cost of this brute-force approach prevents application to
realistic problems. A significant advance was the introduction of the initiator approximation (\mbox{{\it i}-FCIQMC}) \cite{booth1,cleland,cleland11}. In the initiator approximation, only walkers
beyond a certain {\it initiator threshold} $n_{\rm init}$ are allowed to generate walkers on the unsampled determinants. The result is that low-weight determinants whose sign may not be sufficiently accurate, propagate according to a dynamically truncated hamiltonian, defined by the space of instantaneously occupied determinants. This concentrates the stochastic walk within a subspace of the
full Hilbert space allowing for more effective cancellation, at the cost of introducing an initiator error
that may be either positive or negative. However, as the total number of walkers $N_w$ is increased
(for fixed $n_{\rm init}$) the {\it i}-FCIQMC energy converges to the FCI limit.
Additional large efficiency gains can be made by carrying out some of the walk
non-stochastically and by using a multi-determinantal trial wave function when computing the energy estimator,
giving rise to semistochastic quantum Monte Carlo \cite{semistochastic}.
This is not used in the results presented here, but, future studies will investigate the gain in efficiency
and the possible reduction in initiator bias from doing so.

\section{Results and Discussion}
We now describe the application of the DMRG and {{\it i}-FCIQMC} methods
to the beryllium dimer. The beryllium dimer has been of long-standing interest
to theory and experiment. (See Refs. \cite{Patkowski1, Merritt} for an
overview of earlier theoretical and experimental work). Simple molecular orbital arguments
would say that the molecule is unbound, however, Be$_2$ can in fact be observed in the gas phase. The observed bond is
significantly stronger than that of other van der Waal's closed shell diatomics such as He$_2$ and Ne$_2$ \cite{Patkowski1,Patkowski2}.
The unusual bonding arises from  electron correlation effects that are enhanced by the near $sp$ degeneracy of the Be atom.
This near-degeneracy, coupled with the need for very large basis sets to describe the long bond-lengths,
presents a challenge for modern electronic structure methods, while the weak bond makes accurate experimental measurement challenging.
The lack of accurate theoretical data has also hindered the intepretation of experiment, as the
a priori assumed functional form of the potential energy curve biases the extraction of parameters from the spectral lines.
Thus, for many years, there had been significant disagreement between theory and experiment.


The earliest experimental estimate of the well-depth ($D_e$) was 790$\pm$30~cm$^{-1}$ (Ref. \onlinecite{Bondybey, Bondybey2}), but this used a Morse potential
in the fitting that has the wrong shape at large distances, where
van der Waal's forces dominate.
Theoretical calculations generally yielded much deeper wells. Composite
coupled cluster/full-configuration interaction schemes that sum over core/valence (CV), complete basis set (CBS), high-order
correlation effects, and relativistic corrections, gave $D_e$ as 944$\pm 25$~cm$^{-1}$ (Ref. \onlinecite{Martin}),
938$\pm$15~cm$^{-1}$ (Ref. \onlinecite{Patkowski1})
and 935$\pm$10~cm$^{-1}$ (Ref. \onlinecite{koput}). We believe the latter calculation to be the most accurate to date.
Variants of multireference configuration
interaction gave similar, but slightly shallower wells: 903$\pm$8~cm$^{-1}$
(Ref. \onlinecite{Gdanitz1999}, $r_{12}$-MR-ACPF with relativistic corrections),
912~cm$^{-1}$ (Ref. \onlinecite{Schmidt2010}, MRCI with CV, CBS, and relativistic corrections), and 923~cm$^{-1}$ (Ref. \onlinecite{koput}, MRCI+Q, no error bar).
Only recently, remeasurements by Merritt {et al.}\cite{Merritt}, together with an improved fitting of the experimental spectrum,
yielded an experimentally derived $D_e$ consistent with theory: 929.7$\pm$2.0~cm$^{-1}$, which lies  within the  error bars of the  calculations.
A further refit of  Merritt {et al.}'s measurements to a ``fine-tuned'' version of the  potential of Ref. \onlinecite{Patkowski2} gave a slightly modified
well-depth of $D_e=$934.6 cm$^{-1}$, presumably with similar error bars to Ref. \onlinecite{Merritt}. This can be regarded as the most
accurate ``experimental'' estimate of $D_e$ to date.

With the recent resolution of the disagreement between theory and experiment, bonding in the beryllium dimer
can now be considered to be satisfactorily understood, at least
from a computational perspective. Nonetheless, the theoretical efforts so far have required
careful composite schemes  to separately saturate basis set effects, high-order correlation, and core contributions.
While such additive schemes perform quite well, 
{the need to assume additivity between large contributions is theoretically unsatisfactory
and can potentially  introduce some uncertainty into the final predicted result}.
 For example, the all-electron FCI calculation
 in Ref. \onlinecite{Patkowski1} could only be carried out in an aug-cc-pVDZ basis, and gave a well-depth of only 181~cm$^{-1}$, while
the  CCSD(T) calculations in the largest aug-cc-pV7Z basis \cite{koput} gave a well-depth of only 696~cm$^{-1}$.
Thus, in reaching the value of $D_e \approx$ 935~cm$^{-1}$  a large degree of transferability amongst incremental
contributions was assumed. The only non-composite method, the $r_{12}$-MR-ACPF calculation of Gdanitz \cite{Gdanitz1999} gave
a non-relativistic $D_e=898$~cm$^{-1}$, which remains quite far from the best experimental or theoretical results.

\begin{table*}
\caption{\label{tab:dmrgm} Energy in $E_h$ and discarded weights of the DMRG calculation with the canonical transcorrelated Hamiltonian and cc-pCVQZ-F12 basis set for 
the Be$_2$ dimer at a bond length of 2.45 \AA. (l.) and (q.) denote the results of linear and quadratic extrapolations.
}
\begin{tabular}{lcc}
\hline
\hline
M&Energy&Discarded weight\\
\hline
500	&1.57$\times 10^{-7}$	& -29.338592\\
1000&	2.28$\times 10^{-8}$&	-29.338647\\
1500&	5.81$\times 10^{-9}$&	-29.338655\\
2000&	1.56$\times 10^{-9}$&	-29.338657\\
$\infty$(l.)&-&-29.338657\\
$\infty$(q.)&-&-29.338658\\
\hline
\hline
\end{tabular}
\end{table*}

We can now carry out a direct calculation, with saturated large basis sets and explicit correlation as well as a full account of the $n$-electron correlations,
using the canonically transcorrelated DMRG and \mbox{{\it i}-FCIQMC} methods,  thus eliminating the need for
 composite  approaches.
We have computed several points along the ground-state
$1^1\Sigma^+_g$ Be$_2$ potential energy curve using a series of
 cc-pCVnZ-F12 basis sets \cite{basis} with $n$=D, T, Q (henceforth referred to as DZ, TZ, and QZ, for short) and cc-pCVnZ-F12\_OPTRI basis \cite{basis} sets with $n$=D, T, Q respectively for the resolution of the identity (RI) basis sets.
These basis sets contain 68, 124, and 192 basis functions respectively, with
up to $g$ functions in the QZ basis, and the RI basis sets contain 164, 190 and 188 basis functions respectively.
The DMRG calculations
were carried out using the
 \textsc{Block} code \cite{sharma2012spin}.
This DMRG implementation incorporates two symmetries not commonly found
in other implementations: spin-adaptation (SU(2))
and $D_{\infty h}$ symmetries.   Spin-adapted DMRG implementations for
quantum chemistry
were described by Wouters et al. \cite{wouters}
and our group \cite{sharma2012spin},
based on earlier work by McCulloch \cite{mcculloch3}.
Compared to non-spin-adapted DMRG with only $S_z$ symmetry, we find that calculations with $M$ spin-adapted states correspond
in accuracy to approximately $2M$ renormalized non-spin-adapted states in the calculation \cite{sharma2012spin}. Our implementation of $D_{\infty h}$
symmetry resembles that for spin-symmetry, where the Wigner-Eckart theorem is used  to simplify the evaluation of matrix elements as well as to reduce
 storage. We find that $D_{\infty h}$ symmetry brings an additional factor of 2 in the effective $M$ over the use of only $D_{2h}$ symmetry. Consequently, with
both spin and $D_{\infty h}$ adaptation, our reported energies here with $M$ renormalized states are { roughly comparable
in accuracy to similar calculations with $4M$ renormalized states in a conventional DMRG code with only $S_z$ and $D_{2h}$ symmetries}. {Our calculation at the bond length of $2.45\AA$ took a wall clock time of 150 hours running in parallel on 72 Intel Xeon E5-2670 cores, totalling 10,800 core hours.}

\begin{figure*}
\begin{center}
\includegraphics[width=2.5in]{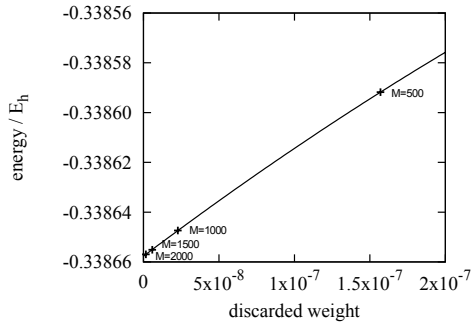}
\end{center}
\caption{Convergence of the DMRG energy (E+29.0) in $E_h$ as a function of the discarded weight
and renormalized states $M$ with the canonical transcorrelated Hamiltonian and cc-pCVQZ-F12 basis set. \label{fig:dmrgm}}
\end{figure*}

The \mbox{{\it i}-FCIQMC} calculations were carried
out using the \textsc{Neci} code \cite{booth1,booth2011,Booth-neci}. These calculations used the Abelian rotational subgroup of $D_{\infty h}$, as described in Ref. \onlinecite{booth2011}. This symmetrized determinant space is smaller than that for the $D_{2h}$ group, especially for large angular momentum basis sets, but is larger than 
that for the full $D_{\infty h}$ group by less than a factor of 2 because it retains only one-dimensional irreducible representations.

The F12 integrals and transcorrelated Hamiltonian were generated
using the \textsc{Orz} code, using the F12 exponent $\gamma=1.0$~$a_0^{-1}$.
The well-depth was calculated from the energy at $r=2.45$~\AA.
All 8 electrons were correlated, thus the largest calculation formally involved more than $3\times 10^{15}$ determinants.
In the DMRG calculations, we also computed the lowest 4 excited states in the $\Sigma$ class of irreps ($2^1\Sigma^+_g$, $1^1\Sigma^+_u$, $1^1\Sigma^-_g$, $1^1\Sigma^-_u$)
at the ground-state equilibrium geometry of $r=2.45\AA$.
For comparison, we also present results of CCSD(T), CCSD(T)-F12, and F12-CCSD(T) \cite{adler2007,knizia,yanai2012canonical} calculations for the ground-state curve,
and MRCI-F12 \cite{shiozaki2011}, MRCI, and EOM-CCSD calculations for the excited states. These computations
were performed using the \textsc{Molpro} package\cite{werner2012}; the
F12-CCSD(T) calculations used the MRCC program with the 
transcorrelated Hamiltonian as input\cite{mrcc}.



\begin{table*}
\caption{\label{tab:ccsdtf12bench}Binding energies in units of m$E_h$ from CCSD(T)/aug-cc-pCVnZ ($n$=4, 5, and 6), F12-CCSD(T)/cc-pCVQZ-F12 and CCSD(T)-F12b/cc-pCVQZ-F12,
as a function of bond length $r$. All the binding energies are counterpoise corrected. 2 different values of the complete basis set limit of the CCSD(T) method are 
calculated by extrapolating the correlation energies of the Be$_2$ dimer and the Be atom (no extrapolation of the HF energy was performed) using Eqs.\ref{extrap1},\ref{extrap3}.}
\begin{tabular}{ccccccccc}
\hline
\hline
&\multicolumn{3}{c}{CCSD(T)}&&\multicolumn{2}{c}{CCSD(T)/CBS}&CCSD(T)-F12b&F12-CCSD(T)\\
\cline{2-4}\cline{6-7}
$r$/\AA&QZ&5Z&6Z&&              $(1)$  &$(2)$      &QZ&QZ\\
\hline
2.20&	0.46&	0.73&	0.86&&	0.97&   1.05&      0.88&	0.86\\
2.40&	2.73&	2.94&	3.04&&	3.12&   3.18&      3.07&	3.05\\
2.45&	2.83&	3.03&	3.12&&	3.20&   3.25&      3.15&	3.14\\
2.50&	2.83&	3.02&	3.11&&	3.18&   3.23&      3.14&	3.12\\
3.00&	1.45&	1.55&	1.60&&	1.64&   1.67&      1.60&	1.61\\
5.00&	0.34&	0.35&	0.36&&	0.36&    0.36&    	0.35&	 0.35\\
\hline
\hline
\end{tabular}
\end{table*}

\begin{table*}
\caption{\label{tab:be2data} Be$_2$ binding energies in units of m$E_h$ as a function of bond-distance
using various methods. The atomic Be energy is \mbox{-14.666740}$E_h$ (DZ-DMRG),
-14.666691$E_h$ (TZ-DMRG), -14.667207$E_h$ (QZ-DMRG). DMRG binding energies for the three basis sets cc-pCVnZ-F12 , where $n$=2, 3, and 4, are tabulated and a fourth column gives our best estimate with error bars (see text for more details). Two sets of {\it i}-FCIQMC calculations are performed, the results in the columns marked QZ(50) and QZ(200) are calculations with 50 million and 200 million walkers respectively. The statistical
error of {\it i}-FCIQMC is denoted in brackets. The difference between the DMRG and {\it i}-FCIQMC numbers is a measure of initiator
error, see text. Merritt, Patkowski denote 
experimentally derived fits from Refs. \cite{Merritt,Patkowski2}.}
\begin{tabular}{cccccccccccc}
\hline
\hline
&\multicolumn{4}{c} {DMRG}&&\multicolumn{2}{c}{{\it i}-FCIQMC}&&\multicolumn{2}{c}{experiment}&\\
\cline{2-5}
\cline{7-8}
\cline{10-11}
r/\AA& DZ        & TZ    &  QZ   & CBS/BSSE/rel.&&  QZ (50) &QZ(200)   && Merritt & Patkowski \\
\hline
2.20 &    0.68    & 1.76   & 2.11   & 2.23(0.08)&&    2.06 (0.02)&&& 2.41   & 2.21 \\
2.30 &    2.33    & 3.41   &    --   &       --       &&--&  &&3.67   & 3.71 \\
2.40 &    3.02    & 3.99   & 4.20   & 4.26(0.05) &&    4.22 (0.02)&&& 4.17   & 4.22 \\
2.45 &    3.13    & 4.09   & 4.24   & 4.30(0.04)&&    4.27 (0.05)&4.21(0.04)&& 4.24   & 4.26 \\
2.50 &    3.13    & 4.00   & 4.18   & 4.24(0.04) &&    4.32 (0.03)&4.11(0.04)&& 4.20   & 4.19 \\
2.60 &    2.95    & 3.70   &    --   & -- &&       --       &&&  3.89   & 3.86 \\
2.70 &    2.64    & 3.27   &    --   &--&&        --      &&&  3.44   & 3.42 \\
3.00 &    1.72    & 2.12   & 2.23 &2.26(0.02)  &&    2.32 (0.05)&&& 2.18   & 2.22 \\
5.00 &    0.31    & 0.37   & 0.39  &0.39(0.01) &&    0.55 (0.03)&&& 0.39   & 0.40 \\
\hline
\hline
\end{tabular}
\end{table*}

\begin{table*}
\caption{\label{tab:des} A comparison of $D_e$ cm$^{-1}$ from this work and
from the literature. Here BSSE, CBS and rel. respectively indicate that corrections have been made for basis set superposition error, basis set incompleteness error and relativistic effects. }
\begin{tabular}{cc}
\hline
\hline
Method         &        \\
\hline
CCSD(T)-F12b/BSSE    & 699.3  \\
DMRG           & 931.2  \\
{\it i}-FCIQMC         & 924(9)       \\
DMRG/CBS/BSSE/rel.& 944(10)     \\
\hline
Author         &       \\
\hline
Merritt(E/T)\cite{Merritt}   & 929.7(2) \\
Patkowski(E/T)\cite{Patkowski2} & 934.6 \\
Patkowski(T)\cite{Patkowski1}   & 938.0(15) \\
Schmidt(T)\cite{Schmidt2010}     & 915.5 \\
Koput(T)\cite{koput}       & 935.1(10) \\
\hline
\hline
\end{tabular}
\end{table*}

Tables \ref{tab:be2data} and \ref{tab:des} present our accumulated data for the DMRG and \mbox{{\it i}-FCIQMC} ground-state Be$_2$ calculations,
as well as selected computed and reference data for the well-depths. All DMRG energies correspond to $M=2000$ (see below) while all \mbox{{\it i}-FCIQMC} calculations
were carried out with $n_{\rm init}=3$ and $N_w=5 \times 10^7$ (see below).
Figure \ref{fig:dmrgm} shows
the convergence of the DMRG energy as a function of the discarded weight and $M$ for the QZ basis at $r=2.45\AA$; {energies as a function of $M$ are given in Table \ref{tab:dmrgm}. We note that the DMRG energies presented in Table~\ref{tab:dmrgm} and Figure~\ref{fig:dmrgm}
were obtained by first carrying out standard DMRG calculations up to $M$=2500, and then {\it backtracking} (by decreasing $M$ in subsequent sweeps) 
down to $M$=500 in steps of 500, to obtain  the tabulated energies at $M$=500, 1000, 1500, 2000. This ensures that the energy at each $M$ is well converged
and free from any initialization bias, leading to more accurate extrapolation.
We calculate the DMRG extrapolated energy by fitting to
linear and quadratic functions of the discarded weight. Due to the high cost of calculation, insufficient 
sweeps were performed at $M$=2500 to attain full convergence, hence the DMRG energies at $M$=2500 were not themselves used in the extrapolation.
The maximum difference between the linear and quadratic extrapolations is 6 $\mu E_h$, and we use this as an upper estimate of the remaining error
in the DMRG energy. Examining Fig. \ref{fig:dmrgm}, we find that the DMRG energy converges extremely rapidly with $M$: even by $M=1000$, the total DMRG energy in the QZ 
basis appears within 10~$\mu E_h$ (2 cm$^{-1}$) of the extrapolated
$M=\infty$ result!

The \mbox{{\it i}-FCIQMC} energies contain two sources of error: statistical error (due to the finite simulation time), and initiator error (due to the
finite walker population). The statistical errors are listed in the Table~\ref{tab:be2data}  and are
on the order of 20-50~$\mu E_h$. The remaining discrepancy between the \mbox{{\it i}-FCIQMC}
energies and the DMRG energies is due to initiator error. Note
that the initiator error can be of either sign. Because of the small energy scales
of this system, the initiator error is significant at some bond-lengths. For
example, at $r=2.5\AA$, the initiator error with $N_w=5\times 10^7$ is 0.14~m$ E_h$, or about $5 \sigma$,
causing the \mbox{{\it i}-FCIQMC} curve to have an unphysical shape (the energy
at $2.50\AA$ is below that at the equilibrium distance $r=2.45\AA$). The initiator error can be removed by carrying out simulations with larger number of walkers. At $r=2.45\AA$ and $r=2.50\AA$ we recomputed the \mbox{{\it i}-FCIQMC} using $N_w=2\times 10^8$ walkers. These \mbox{{\it i}-FCIQMC} are now in better agreement with the 
converged DMRG energies and restore the physical shape of the potential. However, such calculations were 3-4 times more expensive than the corresponding
DMRG calculations.}


We now discuss the possible remaining sources of error and non-optimality in our
calculations. 
These include basis set superposition error (BSSE),  relativistic effects, non-optimality of the F12 $\gamma$ exponent, geometry effects, errors associated with the F12 approximations in the canonical transcorrelation approach and basis set incompleteness error. BSSE error can be estimated from the
counterpoise correction \cite{boys}. We find the counterpoise contribution to
the F12-DMRG well-depth to be  -11~$\mu E_h$ (-2.4 cm$^{-1}$) at the QZ level.
Our relativistic correction using the CCSD(T)/aug-cc-pCVQZ method with the second-order Douglas-Kroll-Hess (DKH) one-electron Hamiltonian is
-4.2 cm$^{-1}$, which is in good agreement with previous studies\cite{Patkowski1,Gdanitz1999}.
We have checked the optimality of the F12 exponent and the bond-length effects
through CCSD(T)-F12 calculations \cite{adler2007,knizia}. At the QZ level, $\gamma=0.8-1.2$ yielded the same CCSD(T)-F12 $D_e$=3.2~m$E_h$ to within 2~$\mu E_h$ (0.4 cm$^{-1}$) and thus we conclude that
our exponent of $\gamma=1.0$  is near-optimal. The difference in energy between the CCSD(T)-F12/QZ equilibrium bond-length energy  (at 2.46\AA),
and the energy at our assumed $r_e=2.45\AA$ is only 3~$\mu E_h$ (0.6 cm$^{-1}$).

{The F12 canonical transcorrelation approach contains two kinds of error. The first is the auxiliary basis integral approximations used
to compute the F12 integrals, and the second is the neglect of normal-ordered three-particle operators in the canonical transcorrelated Hamiltonian
as described above. (We recall that in this work all three-particle cumulants are zero in our definition of $\bar{H}^{F_{12}}$, since we normal order 
with respect to a Hartree-Fock reference). 
Both the above errors are non-variational, which can be seen from the DMRG atomic energies as we increase the basis cardinal number; these are
-14.66674~$E_h$ (DZ), -14.66669~$E_h$ (TZ), -14.66721~$E_h$ (QZ). For comparison, the best variational calculation for the beryllium atom that we are aware of, using exponentially correlated Gaussian expansions, is -14.66736  $E_h$\cite{beatom}. However, both errors also go identically to zero as the orbital basis is increased, because
the F12 factor (and the $A^{F_{12}}$ amplitude) is only used to represent the correlation not captured within the basis set. 

To obtain more insight into the error from the F12 canonical transcorrelated Hamiltonian, we have computed in Table \ref{tab:ccsdtf12bench} the 
 F12-CCSD(T)/cc-pCVQZ-F12 binding energies (i.e. CCSD(T) using the canonical transcorrelated Hamiltonian) using the MRCC
program of K\'{a}llay \cite{mrcc}, and the conventional CCSD(T)-F12b/cc-pCVQZ-F12 binding energies using the \textsc{Molpro} program package \cite{werner2012}. (As pointed out by Knizia et al.\cite{ccsdf12b}, the CCSD(T)-F12b variant is to be preferred with the large basis sets used here).
We observe that the CCSD(T)-F12b/cc-pCVQZ-F12 and F12-CCSD(T)/cc-pCVQZ-F12 binding energies agree very well (to within 5 cm$^{-1}$  along the entire  binding curve). 
Perfect agreement between the methods is not expected as they correspond to different
F12 theories, but these results show that the neglect of three-particle operators in the canonical transcorrelated Hamiltonian produces a description
 with no significant differences
from a standard F12 approach. 

To extrapolate the remaining F12 and basis set errors to zero, we carry out a further basis-set completeness (CBS) study.
In Table \ref{tab:ccsdtf12bench}, we  give the CCSD(T)/aug-cc-pCVnZ binding energies for $n$=4, 5, 6. Following Koput\cite{koput}
we use the following two basis extrapolation formulae to provide error bars on the complete basis result:
 \begin{align}
E_n &= E_{\infty} + a \exp\left( -b(n-2)\right) \label{extrap1}\\
E_n &= E_{\infty} + a /(n+0.5)^b \label{extrap3}
 \end{align}
From Table \ref{tab:ccsdtf12bench} we  observe that the F12-CCSD(T)/cc-pCVQZ-F12 binding energies correspond closely
to those of CCSD(T)/aug-cc-pCV6Z. Using Koput's prescription, we obtain the extrapolated energy as the average of Eqs.~(\ref{extrap1}), (\ref{extrap3}).
At the equilibrium bond length we obtain a basis set limit correction to the DMRG calculation of 87~$\mu E_h$ (19 cm$^{-1}$) and an uncertainty of 43~$\mu E_h$ (10 cm$^{-1}$). (We estimate the uncertainty as half
the extrapolation correction). Thus, the basis-set error remains the largest source of uncertainty in our calculations.

Compared to the experimentally derived well-depths, we find that our directly calculated DMRG (and {\it i}-FCIQMC) well-depths, 931.2 cm$^{-1}$ (924$\pm$ 9 cm$^{-1}$),
are in excellent agreement with the ``experimental'' $D_e$ of 929.7~cm$^{-1}$ (Merritt et al \cite{Merritt})
and 934.6~cm$^{-1}$ (Patkowski et al \cite{Patkowski2}) (Table \ref{tab:des}).
Including the estimated CBS correction (19 cm$^{-1}$), the
counterpoise correction (-2 cm$^{-1}$), and the relativistic correction
(-4~cm$^{-1}$), yields a corrected well-depth of 944~cm$^{-1}$ (DMRG) {with an error estimate of 10 cm$^{-1}$}, which is slightly larger, 
but still in good agreement with the experimental well-depths. Thus, corrected or otherwise, our calculations compare favorably to the very best experimentally derived well-depths to date. 
Compared to CCSD(T)-F12, we find that  quadruples and higher correlations contribute 25\% of the binding energy,
indicating significant correlation effects in the ground-state.

The largest absolute discrepancy between our calculations and the experimentally derived curve appears at the shorter
bond-length of $r=2.20\AA$, where we find the energy (CBS/BSSE/rel. corrected) to be 2.07~m$E_h$ above the equilibrium point as compared to 1.83~m$E_h$ and 2.05~m$E_h$ respectively, in the experimental numbers of Merritt et al. \cite{Merritt} and Patkowski et al. \cite{Patkowski2}. Given the close agreement between
our computations and experiment at all other points on the curve (the agreement between
the corrected DMRG curve with Patkowski's curve is better than 0.05~m$E_h$ at all points) the
discrepancy with Merritt's experimental number is quite large. When measured as a multiple of the theoretical uncertainty, we also find that the largest errors are at $r=2.20\AA$ (2.4$\sigma$) and at $r=3.00\AA$ (4.0$\sigma$). We note that
the inadequacies of Merritt's fit at {\it longer} distances have already been discussed in Ref. \cite{Patkowski2}. Our results further suggest that there are
inaccuracies in Merritt's experimental fit at shorter distances as well.

\begin{table*}
\caption{\label{tab:exc}Low-lying $\Sigma$ excited state energies (in eV) of Be$_2$  calculated using  (F12-)DMRG and the cc-pCVTZ-F12 and cc-pCVQZ-F12 basis sets. The complete
basis set limit and error estimate of the (F12-)DMRG is also given (see text for more details). Excited state energies from the MRCI-F12 and MRCI+Q-F12 methods using the cc-pCVQZ-F12 basis, and the EOM-CCSD method using the cc-pCV5Z basis are also shown.}
\begin{tabular}{ccccccc}
\hline
\hline
\text{State} & \text{DMRG/TZ} &\text{DMRG/QZ}& \text{DMRG/CBS}&\text{MRCI-F12} & \text{MRCI+Q-F12} &\text{EOM-CCSD}\\
\hline
$2^1\Sigma^+_g$ &3.61& 3.59 &3.57(0.02)&3.60&3.54&3.97\\
$1^1\Sigma^+_u$ & 3.58&3.56 &3.55(0.01)&3.70&3.55&3.48\\
$1^1\Sigma^-_g$ & 7.69&7.66 &7.64(0.03)&8.27&8.13&7.33\\
$1^1\Sigma^-_u$ & 4.81&4.78 &4.77(0.02)&4.80&4.75&5.96\\
\hline
\hline
\end{tabular}
\end{table*}

\begin{figure*}
\begin{center}
\subfigure[$2^1\Sigma^+_g$]{\includegraphics[width=2.5in]{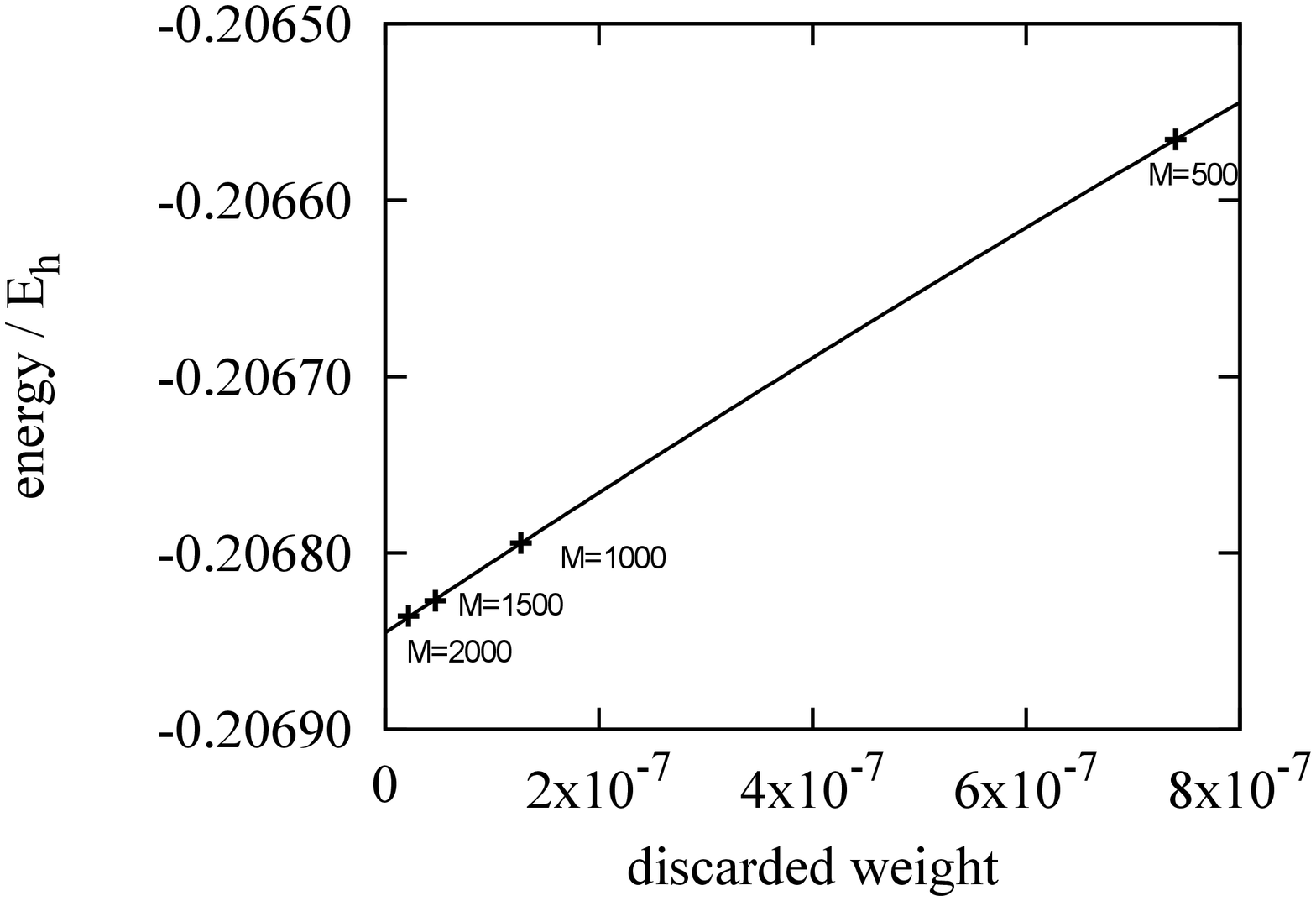}}
\subfigure[$1^1\Sigma^+_u$]{\includegraphics[width=2.5in]{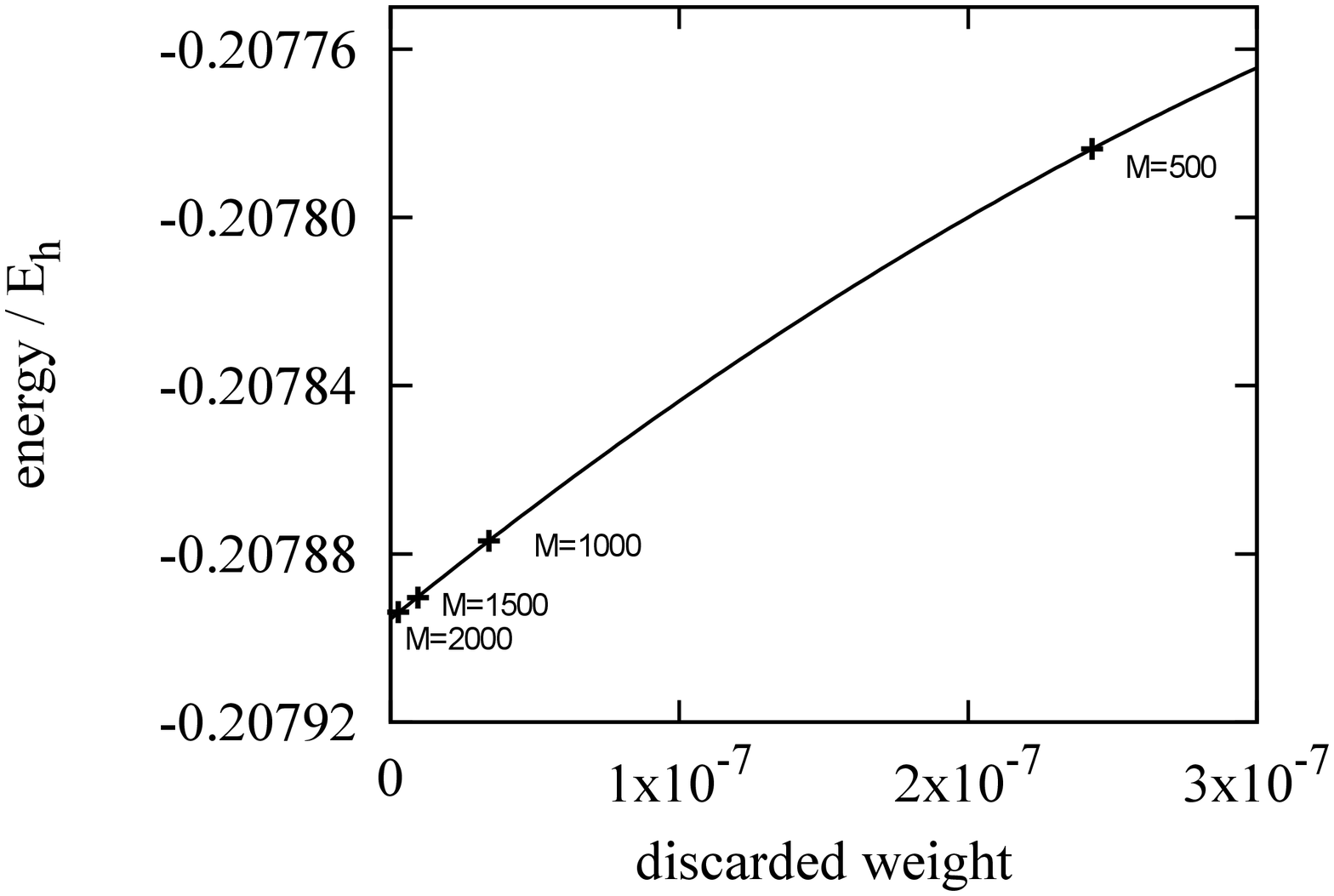}}\\
\subfigure[$1^1\Sigma^-_g$]{\includegraphics[width=2.5in]{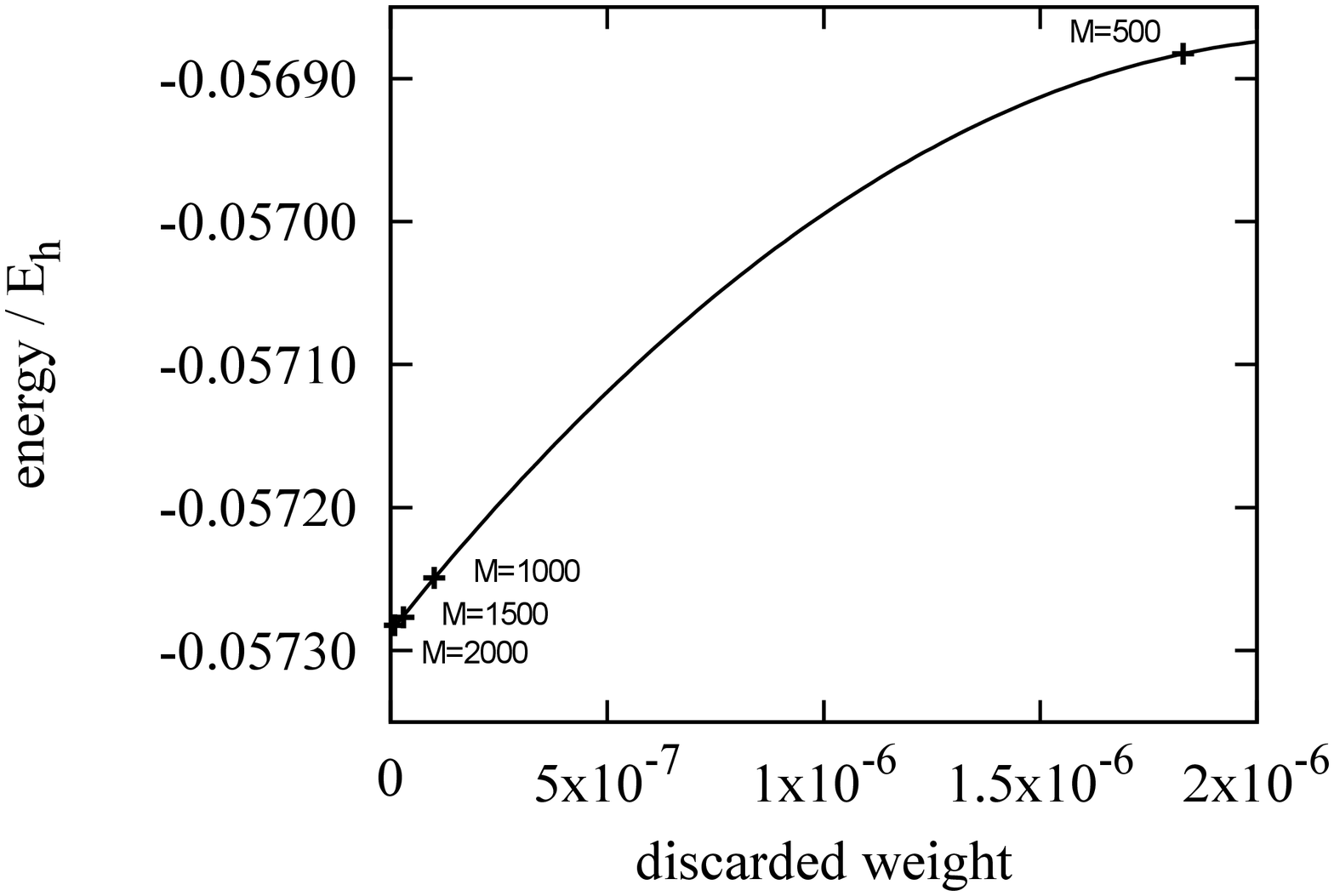}}
\subfigure[$1^1\Sigma^-_u$]{\includegraphics[width=2.5in]{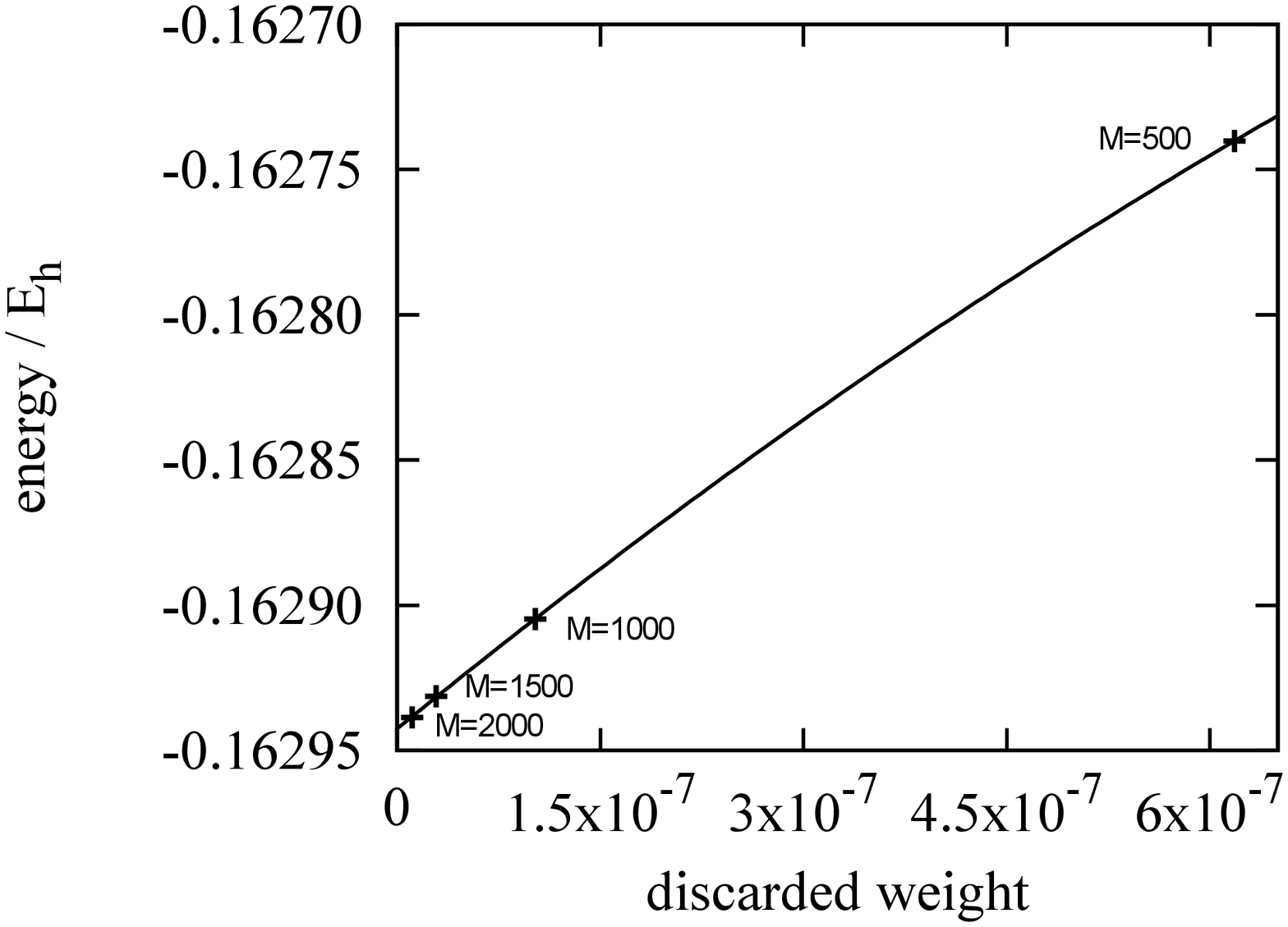}}
\end{center}
\caption{Convergence of the DMRG energies (E+29.0) in $E_h$ for the four excited states as a function of the discarded weight
and renormalized states $M$ with  the canonical transcorrelated Hamiltonian and cc-pCVQZ-F12 basis set. \label{fig:dmrg_excited}}
\end{figure*}

\begin{table*}
\caption{\label{tab:dmrg_excited} Energy in $E_h$ and discarded weights of the DMRG with transcorrelated cc-pCVQZ-F12 basis set on the Be$_2$ dimer at a bond length of 2.45 \AA. 
(l.) and (q.) denote linear and quadratic extrapolations.
}
\begin{tabular}{lcc ccc ccc ccc}
\hline
\hline
&\multicolumn{2}{c}{$2^1\Sigma^+_g$}&&\multicolumn{2}{c}{$1^1\Sigma^+_u$}&&\multicolumn{2}{c}{$1^1\Sigma^-_g$}&&\multicolumn{2}{c}{$1^1\Sigma^-_u$}\\
\cline{2-3}
\cline{5-6}
\cline{8-9}
\cline{11-12}
M&Discarded weight&Energy&&Discarded weight&Energy&&Discarded weight&Energy&&Discarded weight&Energy\\
\hline
500     &7.40$\times 10^{-7}$& -29.206566&&2.43$\times 10^{-7}$&        -29.207784&&1.83$\times 10^{-6}$&  -29.056883     &&6.18$\times 10^{-7}$&  -29.162740\\
1000    &1.27$\times 10^{-7}$&	-29.206794&&3.40$\times 10^{-8}$&	-29.207877&&1.01$\times 10^{-7}$&	-29.057249&&1.02$\times 10^{-7}$&	-29.162905\\
1500    &4.71$\times 10^{-8}$&	-29.206827&&9.48$\times 10^{-9}$&	-29.207890&&2.92$\times 10^{-8}$&	-29.057277&&2.87$\times 10^{-8}$&	-29.162931\\
2000    &2.18$\times 10^{-8}$&	-29.206836&&2.67$\times 10^{-9}$&	-29.207894&&9.74$\times 10^{-9}$&	-29.057282&&1.11$\times 10^{-8}$&	-29.162939\\
$\infty$(l.)&                    &	-29.206844&&                    &	-29.207894&&                    &	-29.057280&&                    &	-29.162940\\
$\infty$(q.)&                    &	-29.206845&&                    &	-29.207895&&                    &	-29.057287&&                    &	-29.162943\\
\hline
\hline
\end{tabular}
\end{table*}

We now turn to the excited state DMRG calculations. While accurate ground-state energies can 
be obtained through composite techniques, this is much more
difficult for excited states, due to significantly larger correlation effects. Near exact excited states, however,
can be accessed through a state-averaged DMRG calculation \cite{Dorando2007}. Combined
with the saturated basis set treatment here, the DMRG excitation energies now allow us to present
very accurate excitation energies for large basis sets, against which other methods
may be compared.
The (F12-)DMRG excitation energies, with comparison MRCI-F12, MRCI+Q-F12, and EOM-CCSD
energies, are shown in Table \ref{tab:exc}.
 The active space used in the MRCI calculations was a 4 electron, 8 orbital complete active space.

The convergence of the DMRG excitation energies with $M$ is shown in Table \ref{tab:dmrg_excited} and is plotted in Figure \ref{fig:dmrg_excited}. 
These show that the DMRG energies are converged to within 10 $\mu E_h$ of the formal exact result, and are thus negligible on the eV scale (on the order of tenths of meV's). The basis set errors for
 the excitation energies are larger than for the ground-state, because we  use the F12 canonical transcorrelated Hamiltonian derived
for the ground $1\Sigma_g^+$ state to compute all the excitation energies,  thus the $A^{F_{12}}$ correlation
factor is biased towards the ground-state. To estimate the complete basis set limit of the excitation energy we use Eq.~(\ref{extrap5}) (derived from a fit to CCSD-F12b energies across a large data set\cite{hill-extrap}).
\begin{align}
E_n &= E_{\infty} + a/n^{4.6} \label{extrap5}
\end{align}
Since excited state complete basis set extrapolation is less well studied, 
we estimate the uncertainty conservatively as twice the difference between the estimated complete basis value and the QZ value. As for the ground-states, the basis set error
remains the largest uncertainty in the calculations, but even
with our conservative estimate ranges only from 0.01 to 0.03 eV.

Overall, MRCI+Q-F12 gives the best agreement with DMRG, with errors of less than 0.05 eV for 3 out of the 4 states. The effect of the Q size-consistency correction 
is significant, contributing as much as 0.15 eV to the excitation energy. The EOM-CCSD excitation energy errors are large for all states,
which is unsurprising given the multireference nature of the ground-state. However, what is most surprising is that
for the $1\Sigma_g^-$ state, the error of the MRCI+Q excitation energy is as large as 0.4 eV! This indicates extremely
strong correlation effects in this state. The $1\Sigma_g^-$ state of the beryllium dimer is thus a good benchmark state for the development of excited state methods.

To summarize, in this work we have used explicit correlation via the canonical transcorrelation approach, in conjunction with
 the density matrix renormalization group and initiator full configuration interaction quantum Monte Carlo methods, to compute the binding curve of the 
beryllium dimer without the
use of composite methods. Our calculations correlate all 8 electrons in basis sets with an orbital basis set of up to 192 basis functions (cc-pCVQZ-F12).
Our direct DMRG calculations produce a well-depth of {$D_e$=931.2 cm$^{-1}$} which
agrees very well with the best experimental and theoretical  estimates.
The remaining basis set effects, BSSE, and relativistic effects, contribute to
a final well-depth of {$D_e$=944 $\pm 10$ cm$^{-1}$}. We find a significant discrepancy between our computed binding energies and the experimentally derived
energies of Merritt et al. at shorter bond-lengths ($r=2.20\AA$) that suggest inaccuracies in the experimental fits. Finally, using DMRG, we have also computed
the excited states at the equilibrium geometry to unprecedented accuracy,  
highlighting surprisingly strong
correlation in the excited states. Overall, we have demonstrated that, by combining explicit correlation with
the DMRG or {\it i}-FCIQMC methods, it is now possible to directly solve the non-relativistic Schr\"odinger equation without significant
basis set or correlation error for small molecules.

\textbf{Acknowledgements}\\
This work was supported by National Science Foundation (NSF) through Grant No. 
NSF-CHE-1265277. TY was supported in part by the Core Research for Grant-in-Aid for Scientific Research (C) (Grant No. 21550027) from Ministry of Education, Culture, Sports, Science and Technology-Japan (MEXT). CJU would like to acknowledge the NSF grant NSF-CHE-1112097 and the Department of Energy (DOE) grant DE-SC0006650.


\end{document}